\documentstyle[preprint,revtex]{aps}
\begin{document}
\draft
\preprint{RU9226}
\begin{title}
Geometry, Scaling and Universality \\
in the Mass Distributions in Heavy Ion Collisions
\end{title}

\author{A.\ Z.\ Mekjian}
\begin{instit}
Deparment of Physics \& Astronomy\\
Rutgers University, Piscataway, New Jersey 08855-0849
\end{instit}
\vspace{-2ex}
\author{S.\ J.\ Lee}
\begin{instit}
 Department of Physics, Kyung Hee University, Suwon 449-701, Korea.
\end{instit}

\receipt{Physical Review C}
\vspace{-2ex}

\begin{abstract}
Various features of the mass yields in heavy ion collisions are
studied. The mass yields are discussed in terms of
iterative one dimensional
discrete maps. These maps are shown to produce orbits for a monomer or
for a nucleus
which generate the mass yields and the distribution of cluster sizes.
Simple Malthusian dynamics and non-linear Verhulst dynamics are
used to illustrate the approach.
Nuclear cobwebbing, attractors of the dynamics, and Lyapanov exponents
are discussed for the mass distribution.
The self-similar property of the Malthusian orbit offers a new variable
for the study of scale invariance using power moments of the mass
distribution. Correlation lengths, exponents and dimensions associated
with scaling relations are developed. Fourier transforms of the mass
distribution are used to obtain power spectra which are investigated
for a $1/f^{\beta}$ behavior.
\end{abstract}


\section{Introduction}

The purpose of this paper is to illustrate how simple
pictures developed from soluble models \cite{mekjian,general,otherx}
can lead to useful methods in
understanding the mass yields in heavy ion collisions.
First, we will attempt to show how iterative one dimensional maps
\cite{popu,hoppen,map}
can characterize ensemble averaged behaviors of these yields.
The language of one dimensional maps contains such ideas as
cobwebbing, orbits and trajectories, attractors, basins of
attraction, Lyapanov exponents, non-linear dynamics and chaos
\cite{general,otherx,popu,hoppen,map,cohen,intm,fractal}.
The approach taken here will also suggest new ways of portraying
the experimental data which go beyond any specific model
discussed here and which may give further insights into the
dynamics of the collision.
One main result of this paper is that the mass yield can be generated
by the orbit of a monomer or of a nucleus (as the initial point)
in its approach to an attractor.
Simple Malthusian dynamics with its associated geometric growth will
be shown to characterize some features of the data considered in this
paper. This geometric Malthusian model will be shown to be an
approximate representation of the soluble models considered in Refs.
[1-3]. A non-linear Verhulst map will also be considered to
investigate signals for non-linear dynamics and limitations to
unbounded geometric growth.

In Refs. [1-3] we have developed an exactly soluble model of fragmentation
and obtained an evolution equation \cite{general} which can be reduced to
a Fokker-Plank equation for a mean cluster distribution.
The evolution equation relates the change of the mean cluster number
of a given size to the mean cluster number of all the other sizes
when the parameter $x$ (introduced in Ref. [1] and discussed below) or
equivalently the temperature changes \cite{general}.
The solution of this evolution equation at a given $x$ value
determines the mean distribution at the corresponding temperature.
We have also shown~\cite{general,otherx} that our $x$ model has a
quite similar structure to a Markov population process~\cite{cohen}.
These properties of our fragmentation model enable us to
relate this model with one dimensional maps for the mean behavior
in a fragmentation process.
The discrete  cluster size in a nuclear fragmentation
corresponds to the iterative index of a discrete map.

Then, secondly, we will discuss the scaling and self-similar behavior
of the power moments \cite{intm} that can be obtained from the mass yield
produced in a nuclear collision. Since a simple linear geometric map
generates orbits having self-similar properties, power moments of the
orbit points also will have a scaling power law behavior. The self similar
property of the geometric or Malthusian model will be shown to offer a
new variable for the study of scale invariance and universality in the
power moments of the mass yield. Moreover, some analytic results for
the scaling properties of the soluble model considered in
Refs.\cite{mekjian,general,otherx}
will be given and compared to the simpler geometric model. The scaling
properties of the power moments with bin size have associated
correlation lengths and exponents related to the dimension of the
system as discussed in Ref. [8] and further developed here.

Thirdly, this paper will investigate the Fourier spectrum of the
fragmentation distribution. The importance of the Fourier spectrum is
the appearance of a power law behavior with frequency $f$, such as
$1/f^{\beta}$ behavior, with $\beta$ an exponent which, in a
description of noise spectra, characterizes the type of
noise\cite{fractal}. For
example $\beta$=0 is white noise, and $\beta$=1, $1/f$ noise. Finally,
we will investigate some experimental data of Ref.\cite{nucdat}.

\section{Geometrical Iterative Maps and Mean Cluster Distributions}

The interrelationships between one dimensional maps and mass yields
developed out of a previous model of fragmentation
and its generalizations~\cite{mekjian,general,otherx}.
Specifically, the model of Refs.\cite{mekjian,general,otherx} contains a
simple expression for the ensemble averaged mean distribution of
cluster sizes $Y_A(k,x) = <n_k>$ given by
\begin{eqnarray}
 <n_k> = \frac{x}{k} \frac{A!}{(A-k)!}
          \frac{\Gamma(x + A - k)}{\Gamma(x + A)} ,  \label{nkx}
\end{eqnarray}
where the $\Gamma$'s are gamma functions.
The $A$ is the total number of nucleons, $k$ the size of the
cluster, and the $x$ is a tuning parameter which can vary from 0
to infinity depending on the temperature.
The parameter $x$ contains the dynamical quantities associated with
the fragmentation.
Thermodynamic arguments~\cite{mekjian,general,otherx} give

\begin{equation}
x ~=~ \frac{V}{v_o} \exp\left[ - \left(\frac{a_v}{k_BT}\right)
             - \left(\frac{k_BT}{\epsilon_o}\right)
               \left(\frac{T_o}{T+To}\right) \right] .
\end{equation}
The $V$ is the freeze out volume, $T$ is the equilibriation
temperature, and $v_o$ is the quantum volume
 $v_o = h^3/(2\pi m_p k_B T)^{3/2}$ with $m_p$
the mass of a nucleon.
The $a_v$ is the coefficient in a simplified equation for the binding
energy $E_B$ of a cluster of $k$ nucleons: $E_B(k) = a_v(k-1)$. The
$\epsilon_o$ is
the level density parameter related to the spacing of excited levels
in a cluster and $T_o$ is a cut-off temperature for internal
excitations. Further details for this relation can be found in Refs.
\cite{mekjian,general,otherx}.
Here, we mention the following. Small $x$ corresponds to a low
temperature  and/or small interaction volume and large $x$ corresponds
to a high temperature and/or large interaction volume. Also for small
$x$, the $<n_k> \sim x$ and thus the $<n_k>$ depend on the separation
energy $a_v$ in the Boltzmann factor $\exp(-a_v/k_BT)$, For large
$x$, $x >> A$, the $<n_k> \sim A (A/x)^{k-1}$ and thus $<n_k> \sim
\exp[a_v(k-1)/k_B T] = \exp[E_B(k)/k_B T]$. In this limit $<n_k>$
depends on the Boltzmann factor in total binding energy.
The $<n_k>$ given by Eq.(\ref{nkx}) thus contains both the
evaporation region dominated by separation energies and the
multifragmentation region dominated by binding energy factors \cite{otherx}.
As a final remark on the physical significance of $x$,
we note that the exponent $- k_BTT_o/\epsilon_o(T+T_o)$,
which also appears as a Boltzmann factor,
is the temperature dependent part of the Helmholtz free energy $F$ per
particle at low temperatures divided by $k_BT$ \cite{general}.
In particular, $F= (3N\epsilon_F/5)(1 - 5 \pi^2(k_BT/\epsilon_F)^2/12)$
where $\epsilon_F$ is the Fermi energy which is related to the level
spacing $\epsilon_o$ via $\epsilon_o = 4 \epsilon_F/\pi^2$.
Note, for low $T$, $T_o/(T+T_o) = 1$ and the dominant term in the factor
$\exp[(-a_v/k_BT)- kT/\epsilon_o]$ is the separation energy part, ie.,
$(-a_v/k_BT)$. For large $T$, the factor $\exp[-k_BT T_o/\epsilon_o
(T+T_o)]$ in $x$ gives rise to the exponential behavior of the
internal partition function of a cluster evaluated in a Fermi gas
approximation. Thus, the factor $\exp[-kT T_o/\epsilon_o(T+T_o)]$
evolves from an evaporation factor into a Boltzmann enhancement factor for
internal excitations. This behavior is similar to the evolution of the
separation energy into a factor for the binding energy of a cluster from
the $\exp(-a_v/k_BT)$ part of $x$. In the rest of this paper, we
will keep $x$ as an unspecified tuning parameter. The above remarks
show that this tuning parameter contains the underlying physical
quantities: $V$, $T$, $\epsilon_o$ and $a_v$.

The simple $x$-model of Eq.(\ref{nkx}) fits the smooth behavior of
the fragment distribution in various proton-nucleus and
nucleus-nucleus collisions~\cite{frgft}. However this model
does not describe all the detailed behavior of these distributions
such as fission peaks and shoulders~\cite{frgft,pronuc}.
One way of overcoming this limitation is using a size $k$
dependent parameter $x_k$ instead of a single parameter $x$.
The size dependence of $x_k$ may arise from a more realistic
treatment of the binding energy by including surface and Coulomb
energies, and shell effects which were not included in obtaining
Eq.(\ref{nkx}).
However, when this is done,
we loose the exact solubability of the model.
Another way of overcoming this limitation is to seek an alternate
methods of approach for fragmentation process. Here, we will show that
the result of Eq.(\ref{nkx}) has an interpretation in terms of iterative
one dimensional maps and use this interrelation as a starting point for
another type of description of the complex dynamics of the
fragmentation process. Iterative maps have been used to model the
behavior of complex systems in many other areas\cite{map}.

The $x$ model of Eq.(\ref{nkx}) has a recursive or iterative property.
Namely, the total average masses in clusters of size $k$, which is
given by $m_k = k <n_k>$, can be related to the total masses of clusters
of size $k + 1$ by
\begin{eqnarray}
 m_{k+1} = r_A(k,x) m_k .   \label{mkmapx}
\end{eqnarray}
Here we dropped the broken bracket representing an ensemble average
of $m_k$ for simplicity in notation.
The recursive coefficient $r_A(k,x)$ is simply
\begin{eqnarray}
 r_A(k,x) = \frac{(A-k)}{(A-k+x-1)} \equiv e^{\lambda_A(k,x)} .  \label{rakx}
\end{eqnarray}
The initial point in the $x$-model is $m_1 = xA/(A+x-1)$.
The mass distribution then follows by successive iterations,
\begin{eqnarray}
 m_1 \to  m_2 \to  m_3 \to \cdots \to  m_k \to \cdots \to  m_A ,  \label{mkseq}
\end{eqnarray}
and satisfies the constraint $A = \sum_k m_k$.

We can also consider the backward version of
the forward invertible mapping given by Eq.(\ref{mkmapx});
\begin{eqnarray}
 m_{A-k} &=& \tilde r_A(k,x) m_{A-k+1} , \nonumber \\
 \tilde r_A(k,x) &=& 1/r_A(A-k,x) .
\end{eqnarray}
Since Eq.(\ref{mkmapx}) is a linear invertible map, this backward map
is the inverse map of the forward map
and gives the sequence of
\begin{eqnarray}
 m_A \to  m_{A-1} \to \cdots \to  m_k \to \cdots \to  m_1 .
\end{eqnarray}

The sequence of Eq.(\ref{mkseq}) gives the orbit of $m_1$
under iteration and the recursive relationship generates
pairs $(m_k, m_{k+1})$
constructed by the recursive coefficient.
This way of representing the results of the above model is similar
to that used in studies of population dynamics and, in particular,
to models in which the population evolves in discrete steps.
The cluster size $k$ serves as the discrete index and corresponds
to the successive generations while the mass distribution $m_k$ is the
analogue of the population size in the $k$-th generation.
The $k$ runs from 1 to $A-1$ and the orbit should be eventually
normalized so that $\sum_{k=1}^A m_k = A$.

In a previous paper~\cite{genetx} another connection of the $x$-model
of fragmentation with a model in population genetics due to
Ewens~\cite{ewens} was explored.
There the index $k$ represented the number of times of a particular
gene appears and $n_k$ the number of different types of genes each
appearing $k$ times in a sample of size $A$.
The parameter $x$ is connected with the mutation rate and effective
population size in population genetics and evolutionary biology.

Using the analogy with population dynamics, the data on the mass
distribution can be plotted as points for each pair ($m_k$, $m_{k+1}$)
for each $k = 1$, ..., $A-1$ with  $m_{k+1}$ along the vertical
axis and $m_k$ along the horizontal axis.
A smooth curve through the data gives the recursion curve
for that data set.
For, example, the simple Malthus model in population dynamics,
which is
\begin{eqnarray}
 m_{k+1} = r m_k = r^k m_1 = r^{k+1} m_0 ,   \label{malthus}
\end{eqnarray}
gives a straight line for $r \ne 1$ in such a plot.
Here $m_0 = m_1/r$ and the constraint $\sum_{k=1}^A m_k = A$
determines $m_1 = A (1-r)/(1-r^A)$.
The parameter $r$ can be determined
by the data set by a least square fit and $r$ represents the slope
of a straight line fit. The parameter $r$ can also be determined by
the slope $\lambda = \log r$ of a straight line fit in a $\log m_k$
{\it vs} $k$ plot.
In general this curve will be more complex than a straight line.
It can depend on $k$ as in Eq.(\ref{rakx}) for the $x$-model
or can depend on $m_k$ itself as in a non-linear map.
The orbit of $m_1$ upon successive iteration is determined by
$m_{k+1} = r^k m_1$
in the Malthus model.

It should be noted that when the $k$ dependence of $r_A(k,x)$
is averaged over then the fragmentation model of Eq.(\ref{mkmapx})
reduces to the simple Malthus model.
For example, a geometric mean $\bar r$ can be defined by
\begin{eqnarray}
 \bar r = e^{\bar\lambda}
    &=& \left[r_A(1,x) r_A(2,x) \cdots r_A(A-1,x)\right]^{1/(A-1)}
            \\
    &=& \exp\left([\lambda_A(1,x) + \lambda_A(2,x) + \cdots
                + \lambda_A(A-1,x)]/(A-1)\right) .
             \nonumber
\end{eqnarray}
The $\bar\lambda$ is the arithmetic mean of
the $\lambda_A(k,x) = \ln r_A(k,x)$.
In the Malthus model or in the fragmentation model of Eq.(\ref{mkmapx})
when $r > 1$ or $x < 1$, $m_k$ increases geometrically with
increasing $k$ and when $r < 1$ or $x > 1$, $m_k$ decreases geometrically
with increasing $k$.
As we shall see, the mass distribution of cluster sizes in
the back direction, i.e., $k$ near $A$, is an important quantity.

Iterative models also lend themselves to a geometrical iteration known
as cobwebbing which involves the generation of the pairs
($m_k$, $m_{k+1}$) by graphical means.
Starting with $m_1$ on the $x$-axis, $m_2$ is evaluated as the
corresponding $y$-axis value on the curve,
$y = r x$, or $m_2 = r m_1$.
Then $m_2$ is reflected onto the $x$-axis through the $45^o$ line,
$y = x$. Next $m_3$ is generated by repeating the procedure
with $m_2$ on the $x$-axis and so on.
The cobwebbing gives a useful visual way of understanding the
trajectory or orbit and gives a large amount of information
by pure geometrical means and simple graphical analysis.

The behavior of $m_k$ as $k$ becomes large ($k$ is limited by $A$ in a
nuclear fragmentation) gives the attractors of the orbits.
In Malthusian dynamics, all orbits of $m$ will
go asymptotically to the $+\infty$ attractor or to the $0$
attractor if $r > 1$ or $r < 1$ respectively.
The point $m = 0$ is an attracting fixed point for $r < 1$
and a repelling fixed point for $r > 1$.
At $r = 1$ and from the result of Eq.(\ref{rakx}) with $x = 1$,
the orbit of $m$ stays fixed for all values of $m$.
For this case, $m_{k+1} = m_k$ forms a fixed line of the map.

Figs.1(a) -- 1(d) illustrate an application of the above ideas
to the experimental data in nuclear fragmentation. From
the fits shown, we see that the fragmentation model
of Eq.(\ref{nkx}) with $x = 0.3$ (solid line) and
its simple Malthusian version (dashed line) give
a reasonably good fit to the data.

These data show that the number of clusters decreases as
the size $k$ increases for small $k$, then slowly
increases until it reaches a somewhat flat region (also called
a ``shoulder'') at large k, then finally
rises sharply again to $k = A$.
The linear map of Eq.(\ref{malthus}) fits the data
better than the $x$-model of Eqs.(\ref{nkx})
and (\ref{mkmapx}) except for the $p + Au$ data of Fig.1(c).
Allowing a simple $k$ dependence in the coefficient
$r$ as $r(k) = a + b k$ with $a = 0.9833$ and $b = 0.000333$,
the linear Malthus map can fit the whole region of the $p + Au$
data (dash-dotted line in Fig.1(c)), (using a $k$ dependent $r$ is
similar to mixing various $r$ values as we will consider in Section VI
for the Au emulsion data~\cite{nucdat}).
This simple $k$ dependent linear map has a geometric mean $\bar r$
of 1.016 and a minimum $n_k$ at $k \approx 85$.
However both the $x$-model and the linear Malthus map do not
produce a levelling off at a large mass region which may be
related to a non-linear map as we will discuss in the next section.

\section{Nonlinear Map}

Non-linear models were introduced to limit the geometric
growth of the Malthus model (attractor at $\infty$).
The classic examples are the Verhulst models;
two examples are
\begin{eqnarray}
 m_{k+1} = \frac{R_V m_k}{m_k + K_V} ,  \label{verhul}
\end{eqnarray}
and the logistic map,
\begin{eqnarray}
 m_{k+1} = R_V m_k (K_V - m_k)    \label{logist}
\end{eqnarray}
with the $K_V - m_k = 0$ if $m_k > K_V$.
The result of Eq.(\ref{logist}) can be rewritten as
$P_{k+1} = a P_k ( 1 - P_k)$.
The $K_V$ is the carrying capacity of the system.
The logistic map Eq.(\ref{logist}) has two fixed points,
$m = 0$ and $m = K_V - 1/R_V$.
Two fixed points of the Verhulst map of Eq.(\ref{verhul})
are $m = 0$ and $m = m_s = R_V - K_V$.
These fixed points of Eqs.(\ref{verhul}) and (\ref{logist})
are all finite in contrast to the Malthus map Eq.(\ref{malthus}).
Since nuclei around $k = 230$ spontaneously fission,
we expect a limit on
the iteration number $k$, $k_{max} \approx 230$,
when we consider nuclear fragmentation.
It should also be noted that in the $x$-model the iteration is
truncated after $k = A-1$ since $r_A(k=A,x) = 0$.

The Verhulst model given by Eq.(\ref{verhul}) can be
solved analytically because the inverse mass $L_k = 1/m_k$
obeys a linear mapping
\begin{eqnarray}
 L_{k+1} = \frac{1}{R_V} + \frac{K_V}{R_V} L_k .
\end{eqnarray}
Due to this linearity, the Verhulst map has a unique inverse map
and thus both the forward map and the backward map generate
the same orbit once the end points ($k = 1$ and $k = A$) are
the same.

The resulting distribution of cluster sizes arising from the orbit
of $m_1 = n_1$ is given by
\begin{eqnarray}
 n_{k+1} = \left(\frac{1}{k+1}\right) \frac{m_s}{(1 - a^k) n_1 + a^k m_s} n_1
                 \label{nkp1}
\end{eqnarray}
where $a = K_V/R_V$ and $m_s = R_V - K_V$. The appearance of $a^k$ in
Eq.(\ref{nkp1}) gives a geometric behavior to $n_{k+1}$.
When $K_V > R_V$, $L_k \to \infty$ and $m_k \to 0$ geometrically.
When $K_V < R_V$, $L_k \to 1/m_s$ and $m_k \to m_s$,
the saturating value (attractor) of the Verhulst mapping of Eq.(\ref{verhul}).
Note $n_k$ falls because of the extra $k$ in $n_k = m_k/k$.
When $R_V = K_V$, then $a=1$ and  $n_{k+1}$  no longer has a geometric part
but rather falls as a power law:

\begin{eqnarray}
 n_{k+1} = \left(\frac{1}{k+1}\right) \frac{1}{1 + n_1 k / R_V} ~ n_1 .
\end{eqnarray}
For large $k$ such that $n_1 k/R_V >> 1$ the $n_{k+1}~ \sim R_V/(k+1)k$.

The $n_k$ has the interesting property of having a bifurcation point
when $a = K_V/R_V = e^{-\zeta} < 1$.
This behavior is not present in $m_k$ which increases uniformly
with $k$ for $a < 1$. The orbit of $m_k$ monotonically approaches
one of the two fixed points (0 or $m_s$) depending on the value of $a$.
The orbit of $n_k$ can have both minimum and maximum points
depending on the value of $a$ and $m_s/m_1$.
The bifurcation point arises from the
extra $k$ in going from $m_k \to n_k = m_k/k$.
Specifically when $m_s/m_1 > e^{2-\zeta} + 1$,
the $n_k$ has a minimum at $k_{min}$ and a maximum at $k_{max}$
given by the solution to the transcendental equation $y_1 = y_2$
with $y_1 = \zeta(k+1) - 1$ and $y_2 = e^{\zeta k} / (m_s/m_1 - 1)$.
The value $k_{min}$ at which $n_k$ is a minimum is bounded from below
by $1/\zeta - 1$ as $m_s/m_1 \to \infty$.
The $k_{1/2}$ point is defined as $m_{k_{1/2} + 1} = m_s/2$
and is given by $k_{1/2} = [\ln(m_s/m_1 - 1)]/\zeta$.
At this point $k_{1/2}$, which is also the inflection point
of $m_k$, the slope $dy_1/dk$ is equal to the slope $dy_2/dk$.
Thus $k_{min} \le k_{1/2} \le k_{max}$
with $k_{min} = k_{1/2} = k_{max}$ when

\begin{eqnarray}
 m_s/m_1 = e^{2 - \zeta} + 1 .  \label{bifurc}
\end{eqnarray}
Thus the condition Eq.(\ref{bifurc}) determines the bifurcation
surface, i.e., the two roots $k_{min}$ and $k_{max}$ merge to
$k_{1/2} = 2/\zeta - 1$,
in the parameter space of ($\zeta$, $m_s$, and $m_1$).

Fig.1(d) also illustrates a fit using the non-linear mapping
of Eq.(\ref{verhul}) to
the $p + Ag$ data where the maximum point $k_{max}$ is placed
near the endpoint $k = A$ to fit the shoulder (dash-dotted curve).
Comparing the result of the $x$-model of Eq.(\ref{nkx}),
the Malthus model, and the Verhulst map near the endpoint
with $k \approx A$, we see some differences.
The result of Eq.(\ref{nkx}) for $x < 1$ approaches the
endpoint with an upward cusp, the Malthus model comes in as a
straight line with positive slope for $r > 1$, and the
Verhulst model can approach the endpoint with zero slope,
positive slope, or negative slope depending on the position
of $k_{max}$ with respect to the endpoint.
The data show a flat region or ``shoulder'' near the endpoint
and then finally a cusp rise.

\section{Self-similarity, Scaling Behavior and Universality
in the\\  \hspace*{0.5in} Power Moments of Mass Distribution}

In this section, we will study the behavior of the simple Malthus
model of Eq.(\ref{malthus}) in greater detail.
The trajectory behavior of a Malthus map can also be seen by looking
at a hypothetical two dimensional spiral
\begin{eqnarray}
 m(r,\theta) = m_1 r^{\theta/2\pi} e^{i\theta} = x + i y , \label{mrthet}
\end{eqnarray}
where $\theta$ can vary from 0 to $\infty$.
The intersection of the continuous spiral with the positive
real axis, i.e., $m(r,\theta) = x_k$ at the values of $\theta = 2\pi(k-1)$,
gives the discrete set of points $x_1 = m_1$ at $\theta = 0$,
$x_2 = m_2 = r m_1$ at $\theta = 2\pi$, and so on.
The index $k$, representing the $k$-th intersection, is a
labeling of clusters by size.
The resulting points become the discrete ``time'' ($t = k$)
version of a continuous system ($t = \theta/2\pi$) in two dimensions.
For $r > 1$ or $x < 1$, the system spirals outward to a fixed point
at infinity and for $r < 1$ or $x > 1$, the system spirals inward to
a fixed point at zero.
For $r = 1$, the spiral reduces to a circle with a radius $m_1$.
In a fragmentation process, $x = 1$ corresponds to the point where the
mass $A$ is uniformly distributed over the cluster index $k$.
The quantities $\lambda_A(k,x)$ which appear in Eq.(\ref{rakx})
can be considered as the Lyapanov exponents for each loop $k$
and $\bar\lambda = \ln \bar r$ the arithmetic mean Lyapanov exponent
of the truncated spiral.

We note that the geometric growth in $m_{k+1}$ of Eq.(\ref{malthus})
or $m(r,\theta)$ of Eq.(\ref{mrthet}) can also be
interpreted in terms of a logarithmic spiral.
Specifically, we consider the logarithmic spiral function
$m(\theta)$ defined by the equation
\begin{eqnarray}
 m(\theta) &=& m_1 e^{\alpha \theta} = e^{\alpha(\theta + \log m_1 /\alpha)} .
      \label{malmap}
\end{eqnarray}
The polar coordinates of the spiral are $(m,\theta)$
and $m_1$ and $\alpha$ are constants.
The mass spectrum $m_1$, $m_2$, ..., $m_k$, ...
corresponds to the radial intercept of the spiral with the $x$-axis.
The successive intercepts give the size index $k$.
Thus $m_2 = m_1 e^{2\pi\alpha}$ and consequently the
coefficient $r = e^{2\pi\alpha}$.
The geometric meaning of $\alpha$ in terms of the logarithmic spiral
can be obtained by considering $\alpha = d(\log m)/d\theta = (d m/d \theta)/m$.
The cotangent of the angle between any radius vector
(the $x$-axis considered as an intercept line here)
and the tangent to the spiral at the intercept point is then
equal to $\alpha$, i.e., $\cot \beta = \alpha$;
the $\beta$ is independent of the orientation of the intercept
radius vector (the $x$-axis here) and the branch of the spiral.
Rotation of the intercept axis (the $x$-axis) by angle of $\theta_0$
corresponds to changes in $m_1$ to $m_1 e^{\alpha\theta_0}$
and not in $r$, $\alpha$, or $\beta$.
Secondly, it should be noted that changes
in $r = e^{2\pi\alpha}$ correspond to changes in $\alpha$
and thus in $\beta$ since $\alpha = \cot \beta$.
As noted in Ref.~\cite{fractal},
the logarithmic spiral is self-similar since $m_1$ can
be absorbed into a constant angle $\log m_1 / \alpha$
as shown in Eq.(\ref{malmap}).

Self-similarity of the geometric sequence 1, $r$, $r^2$, $r^3$, ...
generated by the coefficient $r$ is also discussed in
Schroeder~\cite{fractal}.
Figs.3.10 and 3.11 in Ref.~\cite{fractal} illustrates
a triangular and a rectangular
construction of this property of the infinite series
$1 + r + r^2 + \cdots = 1/(1-r)$ for $r < 1$.
Extending the index $k$ to $-\infty \le k \le \infty$,
the series has an exact self-similar structure.
Thus the power moments of this $m_k$ as considered in Ref.~\cite{intm}
has a power law behavior in the bin size $L$.
However if we restrict the set to be $k = 1$, 2, ..., $A$,
then the power moment would have an $A$ dependence.

In a Malthus map the $q$'th order power moment of the mass distribution
probability $p_k = m_k/A$ with bin size $L$ \cite{intm} is given by
\begin{eqnarray}
 P_q(L,r) &=& \sum_{J=1}^{[A/L]} \left[\sum_{k \in J} p_k\right]^q
           =  \sum_{J=1}^{[A/L]} \left[\sum_{k \in J} r^k m_1/A\right]^q
             \nonumber  \\
          &=& \left[\frac{(1 - r^L)}{(1 - r^A)}\right]^q
              \left[\frac{1 - r^{Aq}}{1 - r^{Lq}}\right] .   \label{pqlr}
\end{eqnarray}
For $r \approx 1$, it is easy to show that $P_q(L,r) = (L/A)^{q-1}$.
Fig.2 shows $P_q(L,r)$ with $r = A/(A+x-1)$.
Fig.3 shows the power moments of the mean mass distribution
\begin{eqnarray}
 \bar{P}_q(L,x) = \sum_{J=1}^{[A/L]} \left[\sum_{k \in J} <p_k>\right]^q
      = \sum_{J=1}^{[A/L]} \left[\sum_{k \in J} \frac{k<n_k>}{A}\right]^q
                     \label{pqxmn}
\end{eqnarray}
in $x$ model of Ref.~\cite{intm}
compared to $P_q(L,r)$ of Eq.(\ref{pqlr}).
Here, the bar over $P_q$ signifies that the power of the mean is
being taken. Below we will also evaluate the mean of the power.
The slope in Fig.3 corresponds to the generalized dimension $D_q$
defined in Ref.~\cite{intm},
\begin{eqnarray}
 D_q(L) = \frac{1}{(q-1)} \frac{\log\left[\bar{P}_q(L,x)\right]}{\log(L/A)} .
                   \label{slopa}
\end{eqnarray}
This figure shows that $D_q$ is about 1 for $x > 1$ and much smaller
than one with $q$ dependence for small $x$.
We can rewrite Eq.(\ref{pqlr}) as
\begin{eqnarray}
 P_q(L,r) = P_q(L,\lambda)
     = \left[\frac{\sinh(\lambda L/2)}{\sinh(\lambda A/2)}\right]^q
       \left[\frac{\sinh(q\lambda A/2)}{\sinh(q\lambda L/2)}\right] ,
             \label{pqlra}
\end{eqnarray}
where $r = e^\lambda$.
This rapidity type quantity $P_q(L,\lambda)$ of $\lambda L$,
which is normalized to 1 at $L = A$, can be
used as a new variable in a study of the power moments.

In the $x$ model, for the bin size $L = 1$ and for $x < 1$,
using Stirling approximation for large $A >> x$,
\begin{eqnarray}
 \Gamma(A+x) \approx (A+x-1)^{x-1} \Gamma(A+1) ,  \label{stirln}
\end{eqnarray}
the power moment $\bar{P}_q(L,x)$ of the mean mass distribution becomes
\begin{eqnarray}
\bar{P}_q(1,x) &=& \sum_{k=1}^A \left[\frac{k <n_k>}{A}\right]^q
           =  \sum_{k=1}^A \left[\frac{x}{A} \frac{A!}{(A-k)!}
               \frac{\Gamma(A+x-k)}{\Gamma(A+x)}\right]^q    \nonumber \\
  & \approx & \left[\frac{x}{A^x}\right]^q
         \left[\left(\Gamma(x)\right)^q + \zeta((1-x)q)\right] ,
              \label{bpqxl1}
\end{eqnarray}
where $\zeta(a)$ is a zeta function.
For $(1-x)q = 1$, the $\zeta((1-x)q)$ should be replaced by
$\log(A-1) + 0.57722$.
Since $\bar{P}_q(A,x) = 1$, then assuming the power moment follow a
power law, we can write $\bar{P}_q(L,x) = (L/A)^{\alpha_q}$ with
$\alpha_q = - \log[\bar{P}_q(1,x)]/\log(A)$ at low $x$. Then

\begin{equation}
\alpha_q ~=~ qx - \frac{q\log[\Gamma(1+x)]
                    + \log[1 + (\Gamma(x))^{-q} \zeta((1-x)q)]}{\log A} .
\end{equation}
Usually, $\alpha_q$ is written as \cite{intm,fractal}
$\alpha_q = D_q (q-1)$ so that the dimension $D_q$ is

\begin{equation}
D_q ~=~ \frac{qx}{q-1} ~- ~ \frac{q\log[\Gamma(1+x)]
   + \log[1 + (\Gamma(x))^{-q} \zeta((1-x)q)]}{(q-1)\log A} .
\end{equation}
Fig.4 show the power moments $\bar{P}_q(L,x)$ for $x =0.5$.
This figure also shows a fit with an $\alpha_q(L) = a_q + b_q/L$
form (thin solid lines) and a fit with the Stirling approximation
of Eq.(\ref{bpqxl1}) for $q = 2$ (thin dash-dotted line).
The $D_q(L)$ of Eq.(\ref{slopa}) are shown in Fig.4(b) for each $q$.

Moreover, for $x=m$ with $m$ an integer, the mean fractional occupancy
in bin $J$ of length $L$ is  given by $\sum k < n_k > /A = \bar
P_J(L,x=m)$, where the
sum $\Sigma$ is over $k$ from $k = (J-1) L+1$ to $JL$. This mean fractional
occupancy can be obtained from

\begin{equation}
\bar{P}_J (L,x=m) ~=~ \frac{(A-1)! m!}{(A+m-1)!} ~ F_J (L,m) .
\end{equation}

\noindent The $F_J (L,m) = H_{J-1} (L,m) - H_J (L,m)$ and,
using Eq.(48) of Ref.\cite{otherx}),
\begin{equation}
H_J (L,m) ~=~ \frac{(m+A_J (L) -1)!}{(A_J (L) -1)! m!} .
\end{equation}

\noindent The $A_J(L) \equiv A-JL$. At $x=m=1$, the $\bar{P}_J(L,1) =
L/A$ and at $x=m=2$, the $\bar{P}_J(L,2)= (2L/A)-(2J-1)(L/A)^2$. The
large $A$ Stirling limit of $\bar{P}_J(L,x)$ is

\begin{eqnarray}
\bar{P}_J (L,x) \cong \left( 1- \frac{(J-1)L}{A} \right)^x - \left( 1-
\frac{JL}{A} \right)^x .
\end{eqnarray}

\noindent With this limit, the power moment $\bar{P}_q(L,x=m)$ can be
rewritten in terms of a simple convolution, namely

\begin{eqnarray}
\bar{P}_q (L,x=m) = \left( \frac{L}{A} \right)^q \sum_{J=1}^{n} ~ (C_J(m))^q
\end{eqnarray}

\noindent with $n = A/L$ and $C_J(m)$ the discrete convolution

\begin{eqnarray}
C_J (m) ~=~ \sum_{k=0}^{m-1}~ b_{J-1}^{m-1-k}~ b_J^k .
\end{eqnarray}

\noindent Here $b_J = 1-(JL/A)$. For $m=1$, $C_J(1) = 1$ and for $m = 2$,
$C_J(2) = (n+1-2J)/n$.
The geometric features in $\bar{P}_q(L,x)$ arise when the factor
$1 - JL/A \simeq \exp (-JL/A)$.
Specifically,

\begin{eqnarray}
 \bar{P}_J(L,x) ~=~ \frac{\exp(-(J-1)Lx/A) - \exp(-JLx/A)}{N(x)}
\end{eqnarray}

\noindent with $N(x) = 1-e^{-x}$ to guarantee that $\bar{P}_J$ is
normalized to unity when summed over $J$. Then,

\begin{eqnarray}
\bar{P}_q(L,x) \equiv \sum_{J=1}^n ~ \bar{P}_J^q(L,x) ~=~
\left[ \frac{{\sinh}(\frac{Lx}{2A})}{{\sinh}(\frac{x}{2})} \right]^q
   ~\left[\frac{{\sinh}(q\frac{x}{2})}{{\sinh}(q\frac{Lx}{2A})}\right] ,
\end{eqnarray}

\noindent which has the same structure of the geometric model of
Eq.(\ref{pqlra}); $\lambda \leftrightarrow x/A$.

Since the Malthus model can describe only the mean behavior of the
mass distribution, we can consider only the power of the mean
distribution, Eq.(\ref{pqlr}).
However for the $x$ model, as we have considered in Ref.~\cite{intm},
we can also study the ensemble average of power moments in each event.
For a small ``integer'' $x$, we can evaluate the power moment,
\begin{eqnarray}
 P_q(L,x) &=& \left<\sum_{J=1}^{[A/L]} \left[\sum_{k \in J} p_k\right]^q\right>
  =  \sum_{J=1}^{[A/L]} \left<\left[\sum_{k \in J} k n_k /A \right]^q\right> ,
\end{eqnarray}
of order 2 analytically.
Using the relations developed in Refs.\cite{general,otherx},
\begin{eqnarray}
 P_2(L,x) &=& \sum_{J=1}^{[A/L]} \left<\left[\sum_{k \in J}
                   k n_k /A \right]^2\right>
           =  \sum_{J=1}^{[A/L]} \sum_{j \in J} \sum_{k \in J}
                   j k <n_j n_k> /A^2               \nonumber  \\
   &=& \sum_{k=1}^A \frac{k^2 <n_k>}{A^2}
       + \sum_{k=1}^{[A/2]} \frac{k^2 <n_k(n_k-1)>}{A^2}
       + \sum_{J=1}^{[A/2L]} \sum_{j\in J} \sum_{k\ne j \in J}
                  \frac{j k <n_j n_k>}{A^2}         \nonumber  \\
   &=& \frac{(A+x)}{A(x+1)}
       + \sum_{J=1}^{[A/2L]} \sum_{j\in J} \sum_{k\in J}
                  \frac{x^2}{A^2} \frac{A!}{(A-j-k)!}
                  \frac{\Gamma(A-j-k+x)}{\Gamma(A+x)} .
\end{eqnarray}
Thus we have
\begin{eqnarray}
 P_2(L,1) &=& \frac{1}{2} \left[\frac{(A+1)}{A} + \frac{L}{A}\right]
                    \nonumber \\
 P_2(L,2) &=& \frac{1}{3} \left[\frac{(A+2)}{A} + 3 \frac{L}{(A+1)}\right]
                    \nonumber \\
 P_2(L,3) &=& \frac{1}{4} \left[\frac{(A+3)}{A}
               + \frac{3(2A^2+3A-1)}{A(A+1)(A+2)} L
               - \frac{3}{A(A+1)(A+2)} L^3 \right] .   \label{pqxi}
\end{eqnarray}
\noindent Eq.(\ref{pqxi}) is valid for bin lengths $L$ which divide $A$
into an ``integer'' number of parts (except $L=A$, thus $L \le A/2$).
These results show a linear dependence of the second order ($q = 2$)
power moments in $L$ rather than a pure power law or exponential
behavior (see Fig.5).
When the $L$ independent part of $P_2(L,m)$ is subtracted out,
the resulting quantity has a
power law behavior in $L$
with an exponent of 1. When the constant part is accepted,
these $P_2 (L,m)$, $m=1$, 2, 3, can be approximated as an exponential.
In Ref.~\cite{intm}, we have shown that the power moments follow
approximately the power law behavior at large $x$ and
exponential behavior at a small value of $x$.
The saturation at large bin size $L$ for large $x$ comes from the high power
dependences in $L$, such as the $L^3$ term in $P_2(L,3)$.

On the other hand, for a bin size of 1,
\begin{eqnarray}
 P_2(1,x) &=& \sum_{k=1}^A \left< \left[\frac{k n_k}{A}\right]^2 \right>
           =  \sum_{k=1}^A \left[\frac{k}{A}\right]^2 <n_k>
            + \sum_{k=1}^A \left[\frac{k}{A}\right]^2 <n_k(n_k-1)>  \nonumber
\\
          &=& \frac{(A+x)}{A(1+x)}
            + \sum_{k=1}^{[A/2]} \left[\frac{x}{A}\right]^2
              \left[ \frac{A!}{(A - 2k)!}
              \frac{\Gamma(A + x - 2k)}{\Gamma(A + x)} \right] .
\end{eqnarray}
Since $<m_k> = k <n_k>$ is small for large $k$ at a large value of $x$,
then using Eq.(\ref{stirln}), the $P_2(1,x)$ can be approximated as
\begin{eqnarray}
 P_2(1,x) & \approx & \frac{(A+x)}{A(1+x)} + \left[\frac{x}{A+x-1}\right]^2
               \left[ \frac{1 - \left(\frac{A}{A+x-1}\right)^A}
                {1 - \left(\frac{A}{A+x-1}\right)^2} \right] .  \nonumber
\end{eqnarray}
For large $A >> x$ with Stirling approximation Eq.(\ref{stirln}),
\begin{eqnarray}
 P_2(1,x) & \approx & \frac{(A+x)}{A(1+x)}
       + \frac{x}{2A} + \frac{x^2 \Gamma(x)}{A^{x+1}} .  \nonumber
\end{eqnarray}
For large $A$ with $x << A$, the leading order $L/A$ dependence
in $P_2(L,x)$ is approximately (c.f. Eq.(\ref{pqxi}))
\begin{eqnarray}
 P_2(L,x) = \frac{1}{x+1} + \frac{x}{A(x+1)} + C_2 \frac{L}{A} , \label{ptlx}
\end{eqnarray}
where
\begin{eqnarray}
 C_2 \approx \frac{x}{2} + \frac{x^2 \Gamma(x)}{A^x} .
\end{eqnarray}
Writing $P_q(L,x) \propto e^{qL/\zeta_q}$ as in Ref.~\cite{intm}, the
correlation length $\zeta_q$ for $q = 2$ is
\begin{eqnarray}
 \zeta_2 = \frac{2(A+x)}{(x+1)C_2} .
\end{eqnarray}
This formula gives value for $\zeta_2$ which approximately agree with
the results of Table 2 in Ref.~\cite{intm}
for small values of $x$.

\section{Fourier Transforms of Mass Distribution
and \\  the $1/\lowercase{f}^\beta$ Power Law}

In this section we propose that Fourier spectra of
fragmentation distributions be studied and we illustrate the
procedure with an example.
The importance of Fourier or power spectrum is the possible
appearance of the ubiquitous $1/f$ noise which appears in
many areas of physics.
Various types of noise can be characterized by writing the
power spectrum associated with the noise as $1/f^\beta$, with $\beta =
0$ for white,
$\beta = 1$ for pink, $\beta = 2$ for brown,
and $\beta > 2$ for black noise.
One mechanism for generating $1/f$ noise is through
intermittency generated by the logistic map near its
tangent bifurcation~\cite{fractal}.
The possible role of the logistic map in nuclear fragmentation
was noted in Section III.
The Malthus model considered in this section will generate a
$1/f^2$ power spectrum.
This example is used for the purpose of illustrating
a powerful tool used in studying properties of discrete
dynamical systems through Fourier analysis.

The dynamical system considered here will be the set of mass
points $m_k$ for $k = 1$, 2, ..., $A$
or its associated probability $p_k = m_k/A$
which are generated by an iterative map given by Eq.(\ref{malthus}).
The Fourier transform of $m_k$ gives the frequency spectrum of the
mass distribution;
\begin{eqnarray}
 {\cal M}_f &=& \sum_{k=1}^A p_k
               \exp\left[- i \frac{2\pi f}{A} k\right]   \nonumber  \\
            &=& \frac{1}{A} \sum_{k=1}^A m_k
               \exp\left[- i \frac{2\pi f}{A} k\right] .   \label{fourt}
\end{eqnarray}
For the mass distribution of Eq.(\ref{malthus}), we have
\begin{eqnarray}
 {\cal M}_f &=& \frac{m_1/r}{A} \sum_{k=1}^A \left[ r
                  e^{- i 2\pi f/A} \right]^k   \nonumber \\
            &=& \frac{m_1}{A} (1 - e^{-i 2\pi f} r^A)
                \frac{e^{-i 2 \pi f/A}}{1 - r e^{- i 2 \pi f/A}} .
\end{eqnarray}
Notice here $(e^{-i2\pi f/A})^A = e^{-i2\pi f} = 1$ for any integer $f$.
An important quantity associated with ${\cal M}_f$ is the
intensity or power spectrum;
\begin{eqnarray}
 I_f = \left| {\cal M}_f \right|^2
    &=& \frac{m_1^2}{A^2} \left[\frac{1 + r^{2A} - 2 r^A \cos(2 \pi f)}
                      {1 + r^2 - 2 r \cos(2 \pi f/A)}\right]   \nonumber \\
    &=& \frac{m_1^2}{A^2} \left(\frac{1 - r^A}{1 - r}\right)^2
        \left[ \frac {1 + \frac{2 r^A}{(1 - r^A)^2}
                 \left(1 - \cos(2\pi f) \right) }
              {1 + \frac{2 r}{(1 - r)^2} \left(1 - \cos(\frac{2\pi f}{A})
                \right) } \right] .      \label{spectr}
\end{eqnarray}
The intensity $I_f$ peaks at half integer $f$ and has a valley at
each integer $f$.
For a Fourier transform in general,
${\cal M}_f$ and $I_f$ are periodic in $f$ with the period of $A$.
Furthermore, $I_f$ has a reflection symmetry with respect to $A/2$,
i.e., $I_{A-f} = I_f$.
Here we consider only integer $f$, i.e., the
valleys of Eq.(\ref{spectr}) for $0 \le f \le A/2$.
At integer $f$, Eq.(\ref{spectr}) reduces to
\begin{eqnarray}
 I_f = \left| {\cal M}_f \right|^2
    &=& \frac{m_1^2}{A^2} \left(\frac{1 - r^A}{1 - r}\right)^2
        \left[1 + \frac{2 r}{(1 - r)^2} \left(1 - \cos(\frac{2\pi f}{A})
                \right) \right]^{-1} .      \label{spectrif}
\end{eqnarray}
For an infinitely large $A$, except for $r = 1$,
$I_f \propto f^0$ for finite $f$ which is a power law in $f$
with the exponent of zero.
This result represent the self-similar property of the Malthus map
discussed in the previous section.
When $r = 1$, which is the same as the $x$ model at $x = 1$,
Eq.(\ref{spectrif}) shows that $I_f = 0$ except for values of $f$
which are integer multiple of $A$.
For $f=A$, $I_A = m_1^2 = 1$ since $m_k = 1$.
For $r$ very small or large compared to one which corresponds to
a large or small value of $x$ in the $x$ model,
the spectrum $I_f$ becomes independent of the frequency $f$ and thus
has a
white power spectrum~\cite{fractal};
if $\epsilon = r - 1 > 2 \pi f/A$, then $I_f$ of Eq.(\ref{spectrif})
exhibits a white spectrum.
On the other hand, for $r$ near unity, the power spectrum
for small $f$ compared to $A$ can be approximated as
$I_f \propto f^{-2}$.
The conditions for a $1/f^2$ power spectrum are, for $\epsilon = r - 1 << 1$,
\begin{eqnarray}
 \epsilon^2 = (r - 1)^2 << \left[\frac{2\pi f}{A}\right]^2 << 1 .
          \label{frange}
\end{eqnarray}
In Fig.1 of Section II, the choice of $r \approx 1.02$
fit the proton-nucleus data for $A \approx 200$.
The $\epsilon A \sim 4$ and the data exhibits $1/f^2$ behavior
for $1 < f < 30$ (see Fig.9(a) in the following Section VI).
The steepest slope occurs at $f = A/4$ and $3A/4$ with
$\pm \left(\frac{m_1}{A}\right)^2 \left(\frac{1 - r^A}{1-r}\right)^2
   \left(\frac{(1 - r)^2}{1 + r^2}\right)^2 \frac{4 \pi r}{A}$.

As a final illustration, we consider the power spectrum associated
with the result of Eq.(\ref{nkx}) at the point $x$=2.
Substituting $m_k = k<n_k>$ into Eq.(\ref{fourt}) we obtain

\begin{eqnarray}
I_f ~=~ \frac{1}{(A+1)^2 ~{\rm sin}^2
(\pi f/A)}
\end{eqnarray}

\noindent for $f=1,2, \ldots A-1$. With low frequencies such
that $\pi f/A << 1$, $I_f \sim 1/\pi^2f^2$ which is a $1/f^{\beta}$
behavior with $\beta=2$.
A more detailed analysis of $1/f^\beta$ power spectrum in mass
distribution will be given in the next section using the Au
emulsion data of Ref.~\cite{nucdat}.

\section{Au Emulsion Data}

In this section we will discuss the Au emulsion data of Ref.~\cite{nucdat}.
The Au emulsion data is an event by event analysis of the charge
distribution $n_z$, with $n_z$ the number of charged $z$ fragments
produced in the collision in each event.
By contrast, the data considered in Fig.1 is an ensemble average
distribution.
Each event conserves the total charge $Z = 79$ event by event in the
Au data.

The ensemble averaged element distribution $<n_z>$ is shown in
Fig.6(a) and the mean charge distribution $p_z = <m_z>/Z$ is
shown in Fig.6(b). These show the strong even-odd oscillation in the
data (filled circles connected with a thin solid line).
Here we mention that the charge distribution $<m_z>$ may be better
than the element distribution $<n_z>$ in distinguishing between
different model. Unlike the results of Fig.1, the Au data cannot be
fit with a single value of $x$ or $r$. The reason for this is the
initial fast fall off with $z$ in $n_z$ or $m_z$.
Fig.6(b) show that two mixed $x$'s in the $x$
model (solid line) or three mixed $r$'s in the Malthus model
(dash-dotted line) give a good fit to the data; both exhibit the fast drop
of $<m_z>$ at small $z$ followed by a flat middle range and then a
rise at the large $z$.
Also shown in Fig.7(a) is the multiplicity distribution $P(M)$
where the multiplicity $M = \sum_z n_z$, event by event.
This figure shows that the emulsion data does not corresponds
to a single $x$ nor a single $r$.

In Refs.\cite{mekjian,otherx,intm}, the importance of the cumulative
mass distribution
\begin{eqnarray}
 M(z) = M_z = \sum_{j=1}^z m_j   \label{cummass}
\end{eqnarray}
was also stressed.
The $M_z$ has a staircase behavior and the sequence
$M_1$, $M_2$, ..., $M_z$, ..., $M_Z$ can be considered as an
ordered sequence similar to the sequence of ``energy'' levels \cite{intm}.
If we use the mean distribution $<m_z>$ in Eq.(\ref{cummass}),
then, at $x = 1$ or $r = 1$, levels are uniformly spaced with
a spacing $s = 1$.
The distribution of the experimental ``spacings'' is then
determined by the experimental values of $m_k = k n_k$ event by event
which may then  be compared with a Poisson or a Wigner distributions.
Fig.7(b) shows the $m_z$ spacing distribution in the Au emulsion
data (solid circle connected by a thin line).
Here $m_z$ is the spacing in each event and the probability is the
ensemble averaged one.
The data looks more like a Wigner distribution ($m_z e^{-m_z/5}/25$)
than a Poisson spacing ($e^{-m_z/10}/10$).
In level spacings, the Wigner distribution is discussed in terms of a
chaotic structure in the spacing distribution.
It should also be noted that the staircase feature \cite{intm} of the
cumulative mass distribution of the data (thin solid line in Fig.8(a))
can be described by two $x$'s (thick solid line) or by
three $r$'s (dash-dotted line).
These three curves are quite similar as can be seen in the figure.

In Fig.8(b) we show a cobweb type plot ($m_z$ {\it vs} $m_{z+1}$)
by solid circles. Also shown in the figure are the pairs of
($m_z$, $m_{z+2}$). The origin of the scatter of the points is
the
even-odd oscillating behavior which also has a varying amplitude in
the charge distribution of Fig.6(b). If the data follows the $x$
model or the Malthus model, then the points would lie on a single
line. The scattered points indicate that even the mean distribution
of the data exhibits some random structure.

Finally, we show the Fourier spectrum of the data in Fig.9.
Fig.9(a) shows the power spectrum of the data compared with
the power spectrum of the $x$ model and the Malthus model.
In the low frequency region, the $x$ model with $x = 0.3$ (dashed line)
exhibits  a $f^{-\beta}$ behavior with $\beta \approx 0.56$,
the Malthus model with a single $r$ (dotted and dash-dot-dot-dotted
lines) has $\beta \approx 1.9$,
and the Malthus model with three mixed $r$'s (dash-dotted line)
follows a
power law with $\beta \approx 0.68$.
The data (solid circle connected by a thin line) shows a more complicated
structure than the $x$ model or the Malthus model. Also shown
by open circles connected with a thin dashed line
is the Fourier spectrum of the data in which the high
peak at $z = 1$ and 2 in the $m_z$ distribution is reduced by putting these
values at the value of $m_3$. This spectrum then follows the single Malthus
model at the low frequency part.
In Fig.9(b), we show the $f$ dependence of the fluctuations of the
data from the mean behavior given by the $x$-model.
The thin dashed line is for $x$ =3, the thick solid line is computed
with 2 $x$'s (66\% $x$ = 0.5 and 34\% $x$ = 50), and the thin solid
line is the Malthus model with three $r$'s. This figure does not
exhibits a simple
$1/f^\beta$ behavior.

\section{Conclusion}

In conclusion, this paper illustrates a simple approach to
the distribution of mass produced in a heavy ion collision.
This distribution can be considered in terms of iterative
one dimensional maps similar to those used in population
dynamics.
In this picture, the orbit of a monomer or nucleon in its approach
to an attractor of the dynamics generates the mass distribution.
Simple Malthusian dynamics seem to account for some features of
the data.
A non-linear model is also discussed to see if the data has any
evidence for non-linearities and limiting behavior in otherwise
geometric growth.
When the Verhulst model is used to generate the mass distribution
$m_k$, the resulting distribution of cluster sizes $n_k = m_k/k$
is shown to have an interesting bifurcation point whose signatures
are discussed.
New ways of representing data which go beyond any specific model
are also suggested.
These ways include plotting pairs ($m_k$, $m_{k+1}$) which
appear in cobwebbing, plotting mass yields, cumulative
mass yields, and inverse mass yields versus cluster size $k$
besides the traditional yields, $n_k = m_k/k$ versus $k$.
More complex maps can also be investigated than
those discussed here
as well as fluctuations such as in Ref.~\cite{intm}.

Using the simple geometric model developed in this paper, questions
related to self-similarity and scaling behavior are discussed. Simple
scaling relations for the power moments of the mass distribution are
developed based on self-similar geometric properties of the mass
yields. The correlation lengths, exponents and dimensions associated
with these scaling relations are discussed. The power spectrum
associated with Fourier transforms of the mass spectrum is also
developed. Under certain conditions a $1/f^{\beta}$ spectrum is found.

One of us (A.M.) would like to thank J.D. Murray of the
University of Washington for his many lectures on methods in
mathematical biology.
He would also like to thank E. Henley and L. Wilets for
the hospitality offered him at the Institute of Theoretical
Physics where this research was started.
This research was supported in part by the N.S.F. grant number
NSF89-03457 and by the DOE.

\figure{Fragment mass distributions in the radiochemical
measurements from the decay of a target residue produced in a collision
of high energy proton on a nucleus.
The filled circles are the data summarized
in Ref.~\cite{pronuc} and the histogram are the thermodynamic statistical
fits of Ref.~\cite{pronuc}.
The solid curve is obtained using the $x$-model of Eq.(\ref{nkx})
with $x = 0.3$ and the dashed line is the orbit under the
Malthus linear iterative map of Eq.(\ref{malthus}).
The coefficients $r$ for the Malthus map are $r = 1.065$
in (a), 1.027 in (b), 1.024 in (c), and 1.041 in (d).
These values are the geometric mean of the corresponding $r_A(k,x)$
with $x = 0.3$ in all four cases. The dash-dotted line in (c) is a
linear map with
the recursive coefficient $r_k = 1 + (k - 50)/3000$.
The dash-dotted line in (d) is an orbit in the non-linear Verhulst
map Eq.(\ref{verhul}) with $\zeta = 0.062$, $m_s = 170$, and $m_1 = 1$.}

\figure{The power moments $P_q(L,r)$ of the mass distribution
in a Malthusian model.
The mapping coefficients $r$ for each $x$ are $r = A/(A + x - 1)$
with $A = 79$.
The solid, dashed, dotted, and dash-dotted lines are for $q = 2$, 3, 4,
and 5 respectively.}

\figure{The power moments $\bar{P}_q(L,x)$ of the mean distribution
in the $x$ model {\it vs}
the power moments $P_q(L,r)$ of the Malthus model of Fig.2 with
the corresponding $x$ values. The lines are same as in Fig.2.}

\figure{Power moments $\bar{P}_q(L,x)$ of the mean distribution of
mass in the $x$ model for $x = 0.5$ and $A = 79$ (a)
and their corresponding slopes, $\alpha_q(L) = (q-1) D_q(L)$
using Eq.(\ref{slopa}) (b).
The solid, dashed, dotted, and dash-dotted
thick lines are for $q = 2$, 3, 4, and 5 respectively. The thin solid
lines in (a), which are close to the thick lines,
are fits using a form  $\bar{P}_q(L) = (L/A)^{(a_q + b_q/L)(q-1)}$
and the thin dashed lines are the simple power law fit
of $\bar{P}_q(L) = (L/A)^{D_q(q-1)}$.
The thin dash-dotted line, which is close to thin dashed line,
is based on Stirling's approximation for $q = 2$, Eq.(\ref{bpqxl1}).
Here, $a_q = D_q - b_q$ with $b_q = 0.04$ and $D_1 = 0.942$, $D_2 = 0.849$,
$D_3 = 0.756$, $D_4 = 0.695$, and $D_5 = 0.656$.}

\figure{Power moments $P_q(L,x)$ which are the ensemble
average of the power of the mass distribution in $x$ model for $A = 79$.
The solid and dashed lines are for $q = 2$ and 3 respectively.
The dotted lines for $x = 1.0$ and 3.0 are the exact results of
Eq.(\ref{pqxi}) and the dotted lines for $x = 0.5$ and 5.0 are
the approximation of Eq.(\ref{ptlx}).}

\figure{The cluster distribution (a) and the charge
distribution (b) of Au emulsion data ($Z = 79$) of Ref.~\cite{nucdat}.
The solid circle connected by a thin line is the data.
The dashed line is the $x$ model with $x = 0.3$ and
the solid line is the $x$ model with a mixture of 0.66 of $x = 0.5$
and 0.34 of $x = 50$.
The dotted line is the Malthus model with $r = Z/(Z+x-1)$ with $x=0.3$ and
the dash-dot-dot-dotted line is for $r$ given by geometrical mean of $x$ model;
the corresponding $r$ values are 1.0089 and 1.0546 respectively.
The dash-dotted line is obtained from a mixture of three $r$'s:
1.1210 (29\%), 0.9938 (40\%), and 0.3492 (31\%) respectively.}

\figure{The multiplicity (a) and the ``level spacing'' (b)
distributions in the ensemble of 415 Au data events
(the solid circle connected by a thin line).
In (a), the dashed line is the $x$ model with $x = 0.5$, the
solid line is the mixture of 0.66 of $x = 0.5$ and 0.34 of $x = 50$,
and the thin solid line is the mixture of 0.72 of $x = 0.5$, 0.15 of $x = 5$,
and 0.13 of $x = 79$.
The dashed and solid lines in (b) are Poissonian ($e^{-m_k/10}/10$)
and Wigner ($m_k e^{-m_k/5}/25$) spacing distribution respectively.}

\figure{Staircase (a) and cobweb (b) plots.
In (a), the curves are the same as in Fig.6 except the thin solid line
is for the data which is very close to the dash-dotted line and to
the thick solid line.
In (b), the solid circles are the pairs of ($m_z$, $m_{z+1}$)
and the closes are the pairs of ($m_z$, $m_{z+2}$).}

\figure{The Fourier transform of the charge distribution $p_z$ (a)
and the noise spectrum (b) for the Au data.
The curves in (a) are the same as in Fig.6.
The open circles connected by a thin dashed line,
are the Au emulsion data except that the charge distribution
at $z = 1$ and 2 are each taken the same as the $z = 3$ value.
This is done to reduce
the effect from the large probability of charge number 1 and 2.
In (b), the thin dashed and the thick solid lines are the noise spectrum
of the Au data with respect to the $x$ model with a single $x$ ($x = 0.3$)
and for two mixed $x$'s (0.66 of $x = 0.5$ and 0.34 of $x = 50$)
respectively. The thin solid line is the noise spectrum with respect
to the Malthus model with three $r$'s as in Fig.6.
Also shown by the solid circles connected with a thin line is the power
spectrum of the data which is shown in (a).}


\begin{thebibliography}{99}
 \bibitem{mekjian}A.Z. Mekjian, Phys. Rev. {\bf C41}, 2103 (1990);
   Phys. Rev. Letts. {\bf 64}, 2125 (1990).
 \bibitem{general}S.J. Lee and A.Z. Mekjian, Phys. Rev. {\bf C45}, 1284 (1992).
 \bibitem{otherx}A.Z. Mekjian and S.J. Lee, Phys. Rev. {\bf A44}, 6294 (1991).
 \bibitem{popu}J.D. Murray, {\bf Mathematical Biology}, Biomathematics
   {\bf V.19}, (Springer-Verlag, 1989).
 \bibitem{hoppen}F.C. Hoppensteadt, {\bf Mathematical Methods of
Population Biology}, (Cambridge University Press, 1982).
 \bibitem{map}R. Devaney, {\bf Chaos and Fractal}, Proceedings of
   Symposia in Applied Mathematics, {\bf V.39}, (R. Devaney and
   L. Keen, eds., American Mathematical Society, 1989).
 \bibitem{cohen} J.E. Cohen, {\it Casual Groups of Monkeys and Man}
   (Harvard University, Cambridge, MA, 1971);  \\
   Theor. Popul. Biol. {\bf 3}, 119 (1972).
 \bibitem{intm}S.J. Lee and A.Z. Mekjian, to be appear in Phys. Rev. {\bf C},
    ``Self-Similarity and Scaling Behavior ...''
 \bibitem{fractal}M. Schroeder, {\it Fractals, Chaos, Power Laws; Minutes
   from an Infinite Paradise}, (W.H. Freeman and Company, New York, 1991).
 \bibitem{nucdat}C.J. Waddington and P.S. Freier, Phys. Rev. {\bf C31}, 888
   (1985).
 \bibitem{frgft}S.J. Lee and A.Z. Mekjian, Phys. Rev. {\bf C45}, 365 (1992).
 \bibitem{pronuc}D.H.E. Gross, L. Satpathy, M. Ta-chung, and M. Satpathy,
   Z. Phys. {\bf A309}, 41 (1982).
 \bibitem{genetx}A.Z. Mekjian, Phys. Rev. {\bf A44}, 8361 (1991).
 \bibitem{ewens}W.J. Ewens, Theoretical Population Biol., {\bf 3}, 87 (1972);
           \\ {\bf Mathematical Population Genetics}, (Springer-Verlag,
   Berlin, 1979).

\end{thebibliography}
\end{document}